\documentclass[showpacs,preprintnumbers,english,prl,twocolumn,amsmath,amssymb]{revtex4}
\usepackage[T1]{fontenc}
\usepackage[latin1]{inputenc}
\usepackage{graphicx}
\usepackage{amssymb}
\usepackage{verbatim}
\usepackage{epic,eepic}
\usepackage{babel}
\begin{document}
\title{Comment on 'Evidence for Stratification of Deuterium-Tritium Fuel in Inertial Confinement Fusion Implosions'}
\author{ Hua Zheng$^{a)}$ and Aldo Bonasera$^{a,b)}$}
\affiliation{
a)Cyclotron Institute, Texas A\&M University, College Station, TX 77843, USA;\\
b)Laboratori Nazionali del Sud, INFN, via Santa Sofia, 62, 95123 Catania, Italy.}
\begin{abstract}
Recent implosion experiments performed at the OMEGA laser facility reported by Casey {\it et al.}\cite{casey},
displayed an anomalously low $dd$ proton yield and a high $tt$ neutron yield as compared to $dt$ fusion reactions, explained
as
a stratification of the fuel in the implosion core. We suggest that in the compression stage the fuel is out of equilibrium. Ions are inward accelerated to a velocity $v_0$ independent on the particle type. 
Yield ratios are simply given by the ratios of fusion cross-sections obtained at the same velocity. 
A 'Hubble' type model gives also a reasonable description of the data. These 
considerations might be relevant for implosion experiments at the National Ignition Facility as well.

\end{abstract}
\maketitle

\begin{figure}
\centering
\includegraphics[width=0.5\columnwidth]{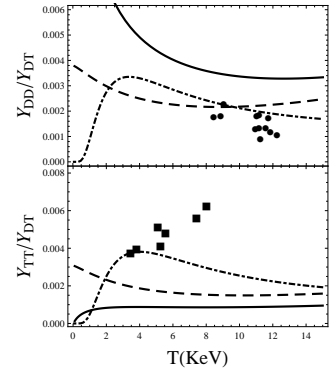}
\caption[]{Yield ratios as function of temperature. Full lines are obtained from reactions rates, Eq. (\ref{rr}), dash-dotted lines from  Eq. (\ref{cs}), and dashed lines from a 'Hubble' type model, Eq. (\ref{density}).}\label{Fig1}
\end{figure}

Assuming thermal equilibrium, the yield ratios of different reactions in imploding-exploding pellets can be obtained as \cite{casey}:
\begin{equation}
\frac{Y_{11}}{Y_{12}} \cong \frac{1}{2}\frac{f_1}{f_2} \frac{\langle \sigma v\rangle_{11}}{\langle \sigma v\rangle_{12}}, \label{rr}
\end{equation}
see \cite{casey} for an explanation of the factors. In Fig. \ref{Fig1}, we plot yield ratios as function of temperature. Eq. (\ref{rr}) overestimates $dd$ fusions (top-panel) and underestimates $tt$ fusions (bottom-panel). A possible explanation given in \cite{casey,amendt} calls for a modification during the process of the ion concentrations, $f_1$, $f_2$,  Eq. (\ref{rr}). While this is surely a possible explanation, we would like to point out that  the fuel is initially cold and at rest. Suddenly, the fuel is accelerated due to the action of the lasers. The fuel is compressed to a maximum density, comes to a halt and finally expands. Thermal equilibrium is not obtained instantaneously and we can assume that,
 in the implosion stage, the fuel is radially accelerated inward to a velocity $v_0$ independent of the particle type. The ratios are now given by:
\begin{equation}
\frac{Y_{11}}{Y_{12}} \cong \frac{1}{2}\frac{f_1}{f_2} \frac{ \sigma(v_0)_{11}}{\sigma(v_0)_{12}}. \label{cs}
\end{equation}
Which can be obtained from Eq. (\ref{rr}) substituting the Maxwellian distribution with a $\delta(v-v_0)$. We can define a 'temperature' as $E=\frac{1}{2}\mu v_0^2=\frac{3}{2}T$ but we stress that the system might not reach  thermal equilibrium; $\mu$ is the $dt$ reduced mass, but it could be the $dd$ or $tt$ reduced mass or an average value, which introduces a small error on the T-axis in Fig. \ref{Fig1}.
The results obtained from Eq. (\ref{cs}) are plotted in Fig. \ref{Fig1}, in much better agreement to data. Assuming a step function in Eq. (\ref{rr}), $\theta(v-v_0)$, gives similar results. 
We have estimated other possible velocity distributions, for instance $v=-\beta r$, a 'Hubble' type implosion \cite{aldo}. If we further assume a Gaussian density distribution at the time when most of the nuclear fusions occur:
\begin{equation}
\rho(r(v))=c \exp(-\frac{r^2}{r_0^2})= c \exp(-\frac{v^2}{v_0^2})=c \exp(-\frac{E}{T}).  \label{density}
\end{equation}
We can define an 'effective temperature' as $T=\frac{1}{2}\mu v_0^2$ and we notice that such a 'temperature' depends on the reduced masses of the reactants.
Thus the yield ratios can be calculated from Eq. (\ref{rr}) with shifted values of $T$ and are plotted in Fig. \ref{Fig1} displaying some agreement to data. This simple example is just to illustrate how a non-equilibrium distribution might look like an equilibrated one. Naturally in the expansion phase thermal equilibrium might be reached and an experimental determination of the plasma distribution is needed.

In conclusion, in this comment we have shown that non-equilibrium effects could explain (part of) the data and their detailed reproduction  might tell us the velocity distribution (and the ion concentrations) at the time when nuclear fusions occur. 

\end{document}